\documentclass[conference]{IEEEtran}
\IEEEoverridecommandlockouts
\usepackage{cite}
\usepackage{amsmath,amssymb,amsfonts}
\usepackage{algorithmic}
\usepackage{textcomp}
\usepackage{tabularx}
\usepackage{graphicx}
\usepackage[table,xcdraw]{xcolor}
\def\BibTeX{{\rm B\kern-.05em{\sc i\kern-.025em b}\kern-.08em\T\kern-.1667em\lower.7ex\ hbox{E}\kern-.125emX}}
\usepackage{hyperref}

\begin{document}

\title{A Comprehensive Review of Leap Motion Controller-based Hand Gesture Datasets}



\author{\IEEEauthorblockN{
Bharatesh Chakravarthi\IEEEauthorrefmark{1},   
Prabhu Prasad B M\IEEEauthorrefmark{2},   
Pavan Kumar B N\IEEEauthorrefmark{3},    
}                                     
\IEEEauthorblockA{\IEEEauthorrefmark{1}
School of Computing and Augmented Intelligence,  Arizona State University, Tempe, Arizona, USA}

\IEEEauthorblockA{\IEEEauthorrefmark{1}
Dept. of Computer Science and Engineering,
Indian Institute of Information Technology,
Dharwad, India }

\IEEEauthorblockA{\IEEEauthorrefmark{2}
Dept.  of Computer Science and Engineering, Indian Institute of Information Technology, Sri City, Andhra Pradesh, India \\ }

*chakravarthi589@gmail.com, 
\IEEEauthorrefmark{2}prabhuprasad1990@gmail.com  \IEEEauthorrefmark{3}pavanbn8@gmail.com }

\maketitle

\begin{abstract}
This paper comprehensively reviews hand gesture datasets based on Ultraleap's leap motion controller, a popular device for capturing and tracking hand gestures in real-time. The aim is to offer researchers and practitioners a valuable resource for developing and evaluating gesture recognition algorithms. The review compares various datasets found in the literature, considering factors such as target domain, dataset size, gesture diversity, subject numbers, and data modality. The strengths and limitations of each dataset are discussed, along with the applications and research areas in which they have been utilized. An experimental evaluation of the leap motion controller 2 device is conducted to assess its capabilities in generating gesture data for various applications, specifically focusing on touchless interactive systems and virtual reality. This review serves as a roadmap for researchers, aiding them in selecting appropriate datasets for their specific gesture recognition tasks and advancing the field of hand gesture recognition using leap motion controller technology.
\end{abstract}

\begin{IEEEkeywords}
Hand gesture, Hand tracking, Leap motion controller, and  Gesture datasets.
\end{IEEEkeywords}

\section{Introduction}
Hand gesture recognition plays a crucial role in facilitating human-computer interaction (HCI), allowing for natural and intuitive communication between humans and machines. This technology has found applications in various domains, including virtual reality, gaming, robotics, and healthcare, where the accurate recognition and interpretation of hand movements and gestures have become increasingly important \cite{b1,b2}. One notable advancement in this field is the Ultraleap's leap motion controller (LMC) \cite{b3}. The LMC is a compact device that can track hand and finger motion with exceptional precision in real-time. It captures even the subtlest nuances of hand gestures, translating them into digital inputs. In comparison to traditional input devices like keyboards and mice, the LMC offers several advantages. By allowing users to interact with virtual and augmented reality environments using their bare hands, the LMC creates an immersive and intuitive experience. Users can reach out, grab objects, manipulate virtual elements, and perform intricate gestures, effectively mirroring their real-world actions. This eliminates the need for handheld controllers or cumbersome input devices, enhancing the sense of presence and reducing the learning curve associated with conventional methods. In essence, the LMC revolutionizes human-computer interaction by providing a seamless and immersive user experience.

\begin{figure}[t]
\centering
\includegraphics[width=1\columnwidth]{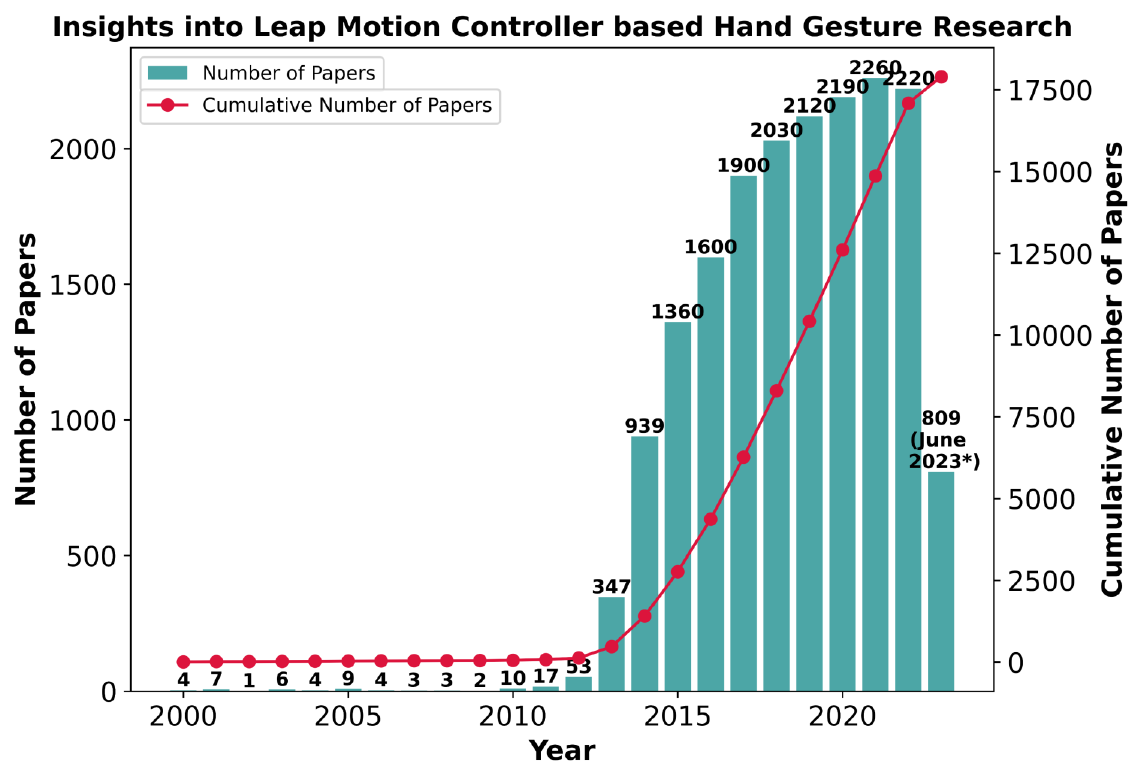}
\caption{Growing trend in the adoption of leap motion controller for hand gesture recognition across various application domains.}
\label{fig01}
\end{figure}

The objective of this review article is to provide a comprehensive overview of hand gesture datasets specifically curated using the LMC. We delve into the technical aspects of the controller, analyzing its hardware and software components that enable precise tracking of hand and finger movements. Furthermore, we explore existing gesture datasets, evaluating their characteristics, diversity, and representativeness in relation to the LMC.
To gain a thorough understanding, we conducted a survey of recent works that have utilized the LMC across different application domains. The results of our exploration, obtained through a search on Google Scholar, reveal a growing trend in the adoption of the LMC for hand gesture recognition, as illustrated in Figure \ref{fig01}. We examined diverse applications, including sign-language recognition, human-computer interaction, virtual reality, robotics, and healthcare, which have utilized gesture datasets acquired from the LMC.
Additionally, we present an experimental evaluation study of Ultraleap's Leap Motion Controller 2 \cite{b3}. This assessment aims to evaluate the device's technical and functional capabilities in generating gesture data for potential application development across various domains.
Furthermore, we highlight the significance of gesture-based touchless interactive systems and explore their potential applications in virtual and augmented reality. By combining the experimental study of the LMC with an exploratory analysis of existing gesture datasets, this review article aims to showcase recent advancements in the field, identify potential applications, address challenges, and inspire future research in the domain of natural and intuitive human-computer interaction.

\begin{figure}[t]
\centering
\includegraphics[width=1\columnwidth]{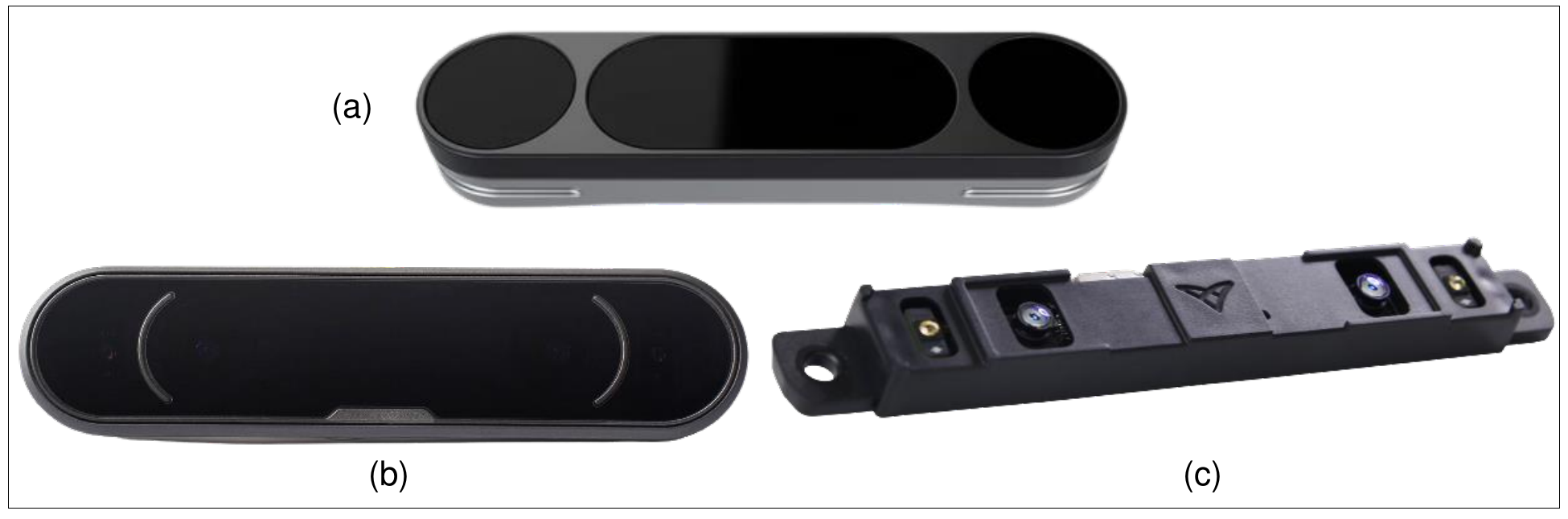}
\caption{Ultraleap's Leap Motion Sensors: (a) Leap Motion Controller 2, (b) Ultraleap 3Di, and (c) Stereo IR 170}
\label{fig02}
\end{figure}

\section{Leap Motion Controller: Technical Overview}

This section presents a comprehensive technical overview of three groundbreaking devices developed by Ultraleap: the Leap Motion Controller 2 \cite{b3}, Ultraleap 3Di Stereo Hand Tracking Camera \cite{b4}, and Stereo IR 170 Camera Module \cite{b5}. These devices have redefined HCI by offering natural and intuitive hand tracking experiences. By exploring their technical specifications, tracking capabilities, and integration possibilities (as summarized in Table \ref{table01}), we can gain valuable insights into their potential applications across diverse industries such as enterprise solutions, gaming, healthcare, education, and robotics.


The Leap Motion Controller 2 (as in Figure \ref{fig02}a) is the next generation of the original LMC, building upon its predecessor's success while introducing advanced features and improved performance. This compact optical hand tracking module offers an enhanced user experience through its smaller form factor, lighter weight, and expanded field of view. By employing advanced optical sensors, it precisely captures the movement of users' hands and fingers within a 3D interactive zone that extends up to 110cm (43") and covers a wide field of view measuring 160° x 160°.
With the integration of Ultraleap's Gemini hand tracking software, the Leap Motion Controller 2 demonstrates the ability to accurately detect and visualize 27 distinct hand elements, including bones and joints, even when they are partially obstructed. This level of precision allows for nuanced tracking of hand gestures, making it well-suited for applications that require precise interaction.
The Leap Motion Controller 2 finds applications in various domains, including AR/VR headsets, PCs, laptops, and desktops. By leveraging this device, users can bring digital content to life through intuitive and gesture-based interactions. It enables seamless and natural interaction with virtual environments, empowering users to manipulate objects and perform actions simply through their hand movements and gestures.

\begin{table}[]
\centering
\caption{Technical Specifications of Leap Motion Sensors}
\label{table01}
\resizebox{\columnwidth}{!}{%
\begin{tabular}{c||c|c|c}
\hline\hline
\textbf{Product} &
  \textbf{Leap   Motion Controller} &
  \textbf{Ultraleap   3Di} &
  \textbf{Stereo IR   170 Evaluation Kit} \\ \hline \hline
\textbf{Dimension}            & 80mm L x 30mm W x 11.3mm H   & 132mm L x 32mm W x 14.5mm H & 145mm L x 18.6mm W x 11.1mm H \\ \hline
\textbf{Weight}               & 29 g                         & 22 g                        & $\sim$30 g                    \\ \hline
\textbf{Data Connection}      & USB type C (3.0)             & USB type-B (2.0)            & USB type-B (2.0 or 3.0)       \\ \hline
\textbf{Tracking Range}       & $\sim$10cm to 60cm(max 80cm) & $\sim$10cm to 75cm(max 1m)  & $\sim$10cm to 75cm(max 1m)    \\ \hline
\textbf{Field of View}        & 140×120°                     & 170×170°                    & 170×170°                      \\ \hline
\textbf{Framerate}            & 115 fps (max)                & 90 fps                      & 90 fps                        \\ \hline
\textbf{Operating Wavelength} & 850 nm                       & 850 nm                      & 850nm                         \\ \hline
\textbf{Public Use}           & Yes                          & Yes                         & No (*as on June 2023)         \\ \hline
\textbf{Usage} &
  Evaluation \& deployment &
  Evaluation \& deployment &
  \begin{tabular}[c]{@{}c@{}}Evaluation \& development \\ prior to integration\end{tabular} \\ \hline
\textbf{\begin{tabular}[c]{@{}c@{}}Operating System \\ Supported\end{tabular}} &
  Windows, macOS, and Android XR2 &
  Windows 10 (min) &
  Windows 7 (min) \\ \hline\hline
\end{tabular}%
}
\end{table}

\begin{table*}[]
\centering
\caption{Comprehensive Summary of Leap Motion Controller based Hand Gesture Datasets }
\label{table02}
\resizebox{\linewidth}{!}{%
\begin{tabular}{c|c|c|c|c|c|c}
\hline \hline
\textbf{Reference} &
  \textbf{Domain} &
  \textbf{\begin{tabular}[c]{@{}c@{}}No. of \\ Gestures\end{tabular}} &
  \textbf{\begin{tabular}[c]{@{}c@{}}Data \\ Modality\end{tabular}} &
  \textbf{Gesture Description} &
  \textbf{Data Collection Setup} &
  \textbf{Key Insights} \\ \hline\hline
{\cite{b6}} &
  \begin{tabular}[c]{@{}c@{}}Language \\ Interpretation\end{tabular} &
  20 &
  \begin{tabular}[c]{@{}c@{}}Image and time \\ series\end{tabular} &
  \begin{tabular}[c]{@{}c@{}}Javanese script characters/letters gestures(/ma, /ga, /ba./ja, /ya, \\ /nya, /la, /pa, /wa etc.,)\end{tabular} &
  Both hands gesture using a single LMC &
  \begin{tabular}[c]{@{}c@{}}Javanese script recognition based on hand gesture, 10 samples \\ from each performer\end{tabular} \\ \hline
{\cite{b7}} &
  \begin{tabular}[c]{@{}c@{}}Human-Robot \\ Interaction\end{tabular} &
  10 &
  Skeletal data &
  \begin{tabular}[c]{@{}c@{}}Military commander sign gestures (Safe, unsafe, hold, stay here, \\ search around etc.,)\end{tabular} &
  Gestures captured at 200 fps using single LMC &
  \begin{tabular}[c]{@{}c@{}}Dynamic hand gestures, three classes of gestures used – less \\ complex. complex, very complex\end{tabular} \\ \hline
{\cite{b8} } &
  \begin{tabular}[c]{@{}c@{}}Activity \\ Recognition\end{tabular} &
  11 &
  \begin{tabular}[c]{@{}c@{}}Image and time \\ series\end{tabular} &
  \begin{tabular}[c]{@{}c@{}}Daily life activities such as utilizing a spoon, fork, and knife, \\ writing, holding a pen, spoon, glass, knife, etc.,\end{tabular} &
  \begin{tabular}[c]{@{}c@{}}7 DoF robotic arm used to hold LMC at an \\ optimal distance\end{tabular} &
  \begin{tabular}[c]{@{}c@{}}Activity data recorded from the dominant arm of the participants \\ while they performed task, 9 subjects with 6 samples of each \\ activity are recorded\end{tabular} \\ \hline
{\cite{b9}} &
  \begin{tabular}[c]{@{}c@{}}User Interface \\ Design\end{tabular} &
  29 &
  Skeletal data &
  \begin{tabular}[c]{@{}c@{}}DICOM viewer and wall screen interaction gestures such as zoom\\ in and out, pan up, down, left and right, next, previous, volume, \\ like, dislike, play/pause, etc.,\end{tabular} &
  \begin{tabular}[c]{@{}c@{}}7 participants performed different tasks using \\ gestures considered in the study with a single  \\ LMC using the QuntumLeap framework\end{tabular} &
  \begin{tabular}[c]{@{}c@{}}Gesture-based interaction demonstration through an image viewer \\ for healthcare workers to browse DICOM medical images and \\ multimedia content on wall screens\end{tabular} \\ \hline
{\cite{b10}} {\cite{b11}} &
  \begin{tabular}[c]{@{}c@{}}Sign-Language \\ Recognition\end{tabular} &
  18 &
  \begin{tabular}[c]{@{}c@{}}Image and \\ skeletal data\end{tabular} &
  \begin{tabular}[c]{@{}c@{}}British and American sign language gestures such as hello/goodbye, \\ you, me, name, sorry, good, bad, time, thank you, etc.,\end{tabular} &
  \begin{tabular}[c]{@{}c@{}}Repeated and differing gestures recorded at a  \\ frequency of 0.2s using LMC and webcam\end{tabular} &
  \begin{tabular}[c]{@{}c@{}}5 subjects contributed 18 gestures for sign language recognition \\ using deep learning algorithms. Dataset publicly available.\end{tabular} \\ \hline
{\cite{b12}} &
  \begin{tabular}[c]{@{}c@{}}Sig-Language \\ Recognition\end{tabular} &
  10 &
  \begin{tabular}[c]{@{}c@{}}Image and \\ skeletal data\end{tabular} &
  \begin{tabular}[c]{@{}c@{}}Russian Fingerspelling signs representing letters such as A, B, C, R, \\ O, T, X, etc.,\end{tabular} &
  \begin{tabular}[c]{@{}c@{}}Sign gestures performed by participants (20 \\ times each) using a single LMC under the \\ supervision of a language interpreter\end{tabular} &
  \begin{tabular}[c]{@{}c@{}}Consists of an overall 800 Russian fingerspelling signs collectively, \\ a Deep learning-based sign language recognition system with \\ 97.5\% accuracy\end{tabular} \\ \hline
{\cite{b13}} {\cite{b14}} &
  \begin{tabular}[c]{@{}c@{}}Human-Computer \\ Interaction\end{tabular} &
  11 &
  Skeletal data &
  \begin{tabular}[c]{@{}c@{}}DICOM interaction gestures such as thumb taps, left and right rotate, \\ release up/down, grip in/out, swipe right/left, etc.,\end{tabular} &
  \begin{tabular}[c]{@{}c@{}}Single LMC used with a laptop. 120 different \\ participants repeating the same action 5 times\end{tabular} &
  \begin{tabular}[c]{@{}c@{}}The dataset is publicly available with over 6600 samples. A gesture \\ recognition system for touch free surgical procedures with 93\% \\ accuracy\end{tabular} \\ \hline
{\cite{b15}} &
  \begin{tabular}[c]{@{}c@{}}Augmented \\ Reality\end{tabular} &
  5 &
  \begin{tabular}[c]{@{}c@{}}Image and \\ skeletal data\end{tabular} &
  Gestures representing victory, OK, thumbs up, loser and call me. &
  Single LMC used with a MacOS PC &
  \begin{tabular}[c]{@{}c@{}}1600 samples with an average of 320 images for each class of \\ gestures. A deep learning-based system with  96\% recognition rate.\end{tabular} \\ \hline
{\cite{b16}} &
  \begin{tabular}[c]{@{}c@{}}Sign-Language \\ Recognition\end{tabular} &
  44 &
  Skeletal data &
  \begin{tabular}[c]{@{}c@{}}Arabic sign language gestures indicating asked, catch, take,  push, \\ write, yesterday, tomorrow, south, west, seat, etc.,\end{tabular} &
  Single LMC used with a PC &
  \begin{tabular}[c]{@{}c@{}}5 different signers performed 44 gestures (29 single-hand, 15 two\\ hand) with over 10 samples for each class\end{tabular} \\ \hline
{\cite{b17}} &
  \begin{tabular}[c]{@{}c@{}}Sign-Language \\ Recognition\end{tabular} &
  22 &
  \begin{tabular}[c]{@{}c@{}}Image and \\ time series\end{tabular} &
  \begin{tabular}[c]{@{}c@{}}Sign language gestures such as poke, pull, slap, press, cut, like, \\ sorry, etc.,\end{tabular} &
  Single LMC used with a PC &
  An overall 840 samples of data \\ \hline
{\cite{b18}}{[\cite{b19}} &
  \begin{tabular}[c]{@{}c@{}}Automotive \\ Interaction\end{tabular} &
  12 &
  \begin{tabular}[c]{@{}c@{}}Image and \\ skeletal\end{tabular} &
  \begin{tabular}[c]{@{}c@{}}Interactive gestures representing actions such as pinch, flip-over, \\ right/left – top/down swipe, clockwise/anticlockwise  rotations, etc.,\end{tabular} &
  \begin{tabular}[c]{@{}c@{}}Dynamic hand gestures acquired in a real car \\ dashboard using LMC and other sensors\end{tabular} &
  \begin{tabular}[c]{@{}c@{}}The dataset is publicly available 3D CNN  based gesture \\ classification system with 94.4\% accuracy\end{tabular} \\ \hline
{\cite{b20}} &
  \begin{tabular}[c]{@{}c@{}}Gesture \\ Recognition\end{tabular} &
  13 &
  Skeletal data &
  \begin{tabular}[c]{@{}c@{}}Gestures include pointing, catching, shaking, zooming, scrolling, \\ rotating, slicing, etc.,\end{tabular} &
  \begin{tabular}[c]{@{}c@{}}Dynamic hand gestures captured using single \\ LMC\end{tabular} &
  \begin{tabular}[c]{@{}c@{}}21 participants performed 50 gesture sequences resulting in 608 \\ instances. HIF3D features based hand trajectories\end{tabular} \\ \hline
{\cite{b21}} &
  \begin{tabular}[c]{@{}c@{}}Touchless \\ Interaction\end{tabular} &
  11 & Skeletal data
   &
  \begin{tabular}[c]{@{}c@{}}Gestures include rotation, increase/decrease contrast, moving left/\\ right/next/previous, zoom in/out, etc.,\end{tabular} &
  \begin{tabular}[c]{@{}c@{}}3D dynamic hand gestures collected using \\ single LMC\end{tabular} &
  \begin{tabular}[c]{@{}c@{}}Gestures performed by 10 different subjects with 5 times repetitions  \\ resulting in 550 samples. The dataset is publicly available\end{tabular} \\ \hline \hline
\end{tabular}%
}
\end{table*}

The Ultraleap 3Di Stereo Hand Tracking Camera revolutionizes interactive screens, seamlessly transforming them into touchless and immersive surfaces. Purposefully designed, this camera (as shown in Figure \ref{fig02}b) enables precise hand tracking in three dimensions, eliminating the need for physical touch. It is designed to perform well even in challenging lighting conditions, along with Gemini hand tracking software module. With its compact dimensions of 132mm L x 32mm W x 14.5mm H and weighing just 22g, the 3Di camera seamlessly integrates with various devices via USB 2.0. Its wide field of view measuring 170° x 170° captures comprehensive hand movements and gestures, providing a truly immersive interaction experience. The camera's impressive tracking range extends from approximately 10cm to 75cm, with a maximum range of up to 1 meter, accommodating various user positions and distances. Designed for reliability, the ruggedized 3Di camera excels in permanent installations across industries such as self-serve kiosks, digital out-of-home installations, retail displays, museums, theme parks, and medical/industrial environments. As a key component of Ultraleap's TouchFree solution, it enables touchless interaction experiences, enhancing safety, hygiene, and user engagement.

The Ultraleap Stereo IR 170 Camera Module (as in Figure \ref{fig02}c) represents a significant leap forward in optical hand tracking technology, revolutionizing the way we interact with digital content. This advanced module captures precise hand and finger movements, enabling seamless and intuitive interaction experiences. Designed to integrate seamlessly into enterprise-grade hardware solutions, displays, installations, and virtual/augmented reality headsets, it serves as a valuable tool for prototyping, research, and development in the AR/VR/XR space. Compared to its predecessor, the Stereo IR 170 Camera Module offers remarkable enhancements. It boasts an extended tracking range, allowing it to accurately track hands within a 3D interactive zone ranging from 10cm (4") to 75cm (29.5") or even beyond. With a wider field of view, users can enjoy a more immersive and encompassing interaction experience. Additionally, the module exhibits lower power consumption and a smaller form factor, ensuring optimal performance without compromising on convenience. From touchless public interfaces and captivating entertainment experiences to healthcare, therapy, education, personnel training, industrial design, engineering, robotics, and remote collaboration, its seamless integration options make it a go-to solution for diverse use cases.


\section{Hand Gesture Datasets}

The leap motion controller sensor has garnered significant attention in the field of hand gesture recognition, captivating researchers from various domains. Its capabilities have led to advancements in language interpretation, human-robot interaction, activity classification, and user interface design. In this section, we explore noteworthy research papers that have utilized the LMC to develop comprehensive hand gesture datasets and delve into gesture recognition and interpretation. Table \ref{table02} provides a summary of the LMC-based hand gesture datasets used across different domains. Let's take a closer look at these studies

Nasir et al. \cite{b6} focused on preserving cultural heritage through hand gesture recognition using the LMC. Their recognition system detects and interprets hand gestures associated with the ancient Javanese script, achieving an average accuracy rate of 95\%. This research demonstrates the potential of the LMC in safeguarding and promoting cultural traditions.
Sesli et al. \cite{b7} explored hand gestures for human-robot interaction in the military domain. They proposed a deep neural network approach using the LMC to analyze complex gestures performed with dynamic movements. This method achieved an impressive accuracy rate of 88.44\% and highlighted the LMC's potential in enhancing communication between humans and robots in military settings.
Sharif et al. \cite{b8} made a contribution to hand gesture recognition, specifically in recognizing activities of daily living (ADL). By combining time and frequency domain features with advanced classifiers, they achieved a classification accuracy rate of over 99\%. This research underscores the precise capabilities of the LMC in recognizing and understanding daily activities based on hand gestures.

Sluÿters et al. \cite{b9} present QuantumLeap, a framework for gestural user interfaces using the LMC. Their modular pipeline addresses programming challenges by acquiring, segmenting, recognizing, and mapping gestures to application functions. The authors validate QuantumLeap through two gesture-based applications, showcasing its practicality and usability. This research contributes insights into gesture-driven systems using the LMC. In another study, Bird et al. \cite{b10} demonstrate the potential of combining computer vision with LMC data for sign language recognition. Their multimodal approach achieves improved recognition accuracy (94.44\%) by combining image and LMC data classification. The study highlights transfer learning from British Sign Language (BSL) to American Sign Language (ASL). This research leverages the LMC's ability to facilitate effective sign language interpretation. Enikeev et al. \cite{b12} conducted research on Russian fingerspelling recognition, specifically targeting the distinct component of sign language. They developed a system that utilizes the LMC to recognize static gestures of Russian fingerspelling. Through deep learning techniques and a dataset comprising 400 static images, their system achieved an impressive success rate of 97.5\%. This research makes a noteworthy contribution to the recognition of fingerspelling gestures in Russian Sign Language, enhancing communication accuracy and efficiency for individuals who rely on this language.

Ameur et al. \cite{b13} conducted a study to address the need for hand gesture datasets tailored specifically to dynamic gestures in surgical procedures. They introduced the LeapGestureDB dataset \cite{b14}, consisting of 6,600 samples of 11 distinct dynamic hand gestures performed by over 120 participants. The authors proposed a 3D dynamic hand gesture recognition approach, achieving an accuracy of 93\%. This dataset significantly contributes to research in dynamic hand gesture recognition, particularly in surgical settings where precise and accurate gesture recognition is crucial. Huo et al. \cite{b15} developed an interactive user interface that integrates augmented reality (AR) and the LMC for hand gesture recognition and translation. Their research investigates the effectiveness of various machine-learning approaches in recognizing five specific hand gestures. The experimental results emphasize the practicality of real-time text translation using hand gesture recognition, thereby reducing communication barriers for individuals with hearing impairments and fostering inclusive communication environments.

Alnahhas et al. \cite{b16} introduces an innovative method for recognizing words in Arabic Sign Language using the LMC. Their approach involves constructing a 3D model of the human hand through infrared technology and extracting features based on the chronology of successive frames. By utilizing a recurrent neural network, their method achieves remarkable classification rates of 89\% for one-hand gestures and 96\% for two-hand gestures. This research significantly contributes to the advancement of Arabic Sign Language recognition, enhancing communication accessibility for individuals who rely on this expressive form. Yang et al. \cite{b17} propose a two-layer Bidirectional Recurrent Neural Network for dynamic hand gesture recognition using the LMC. Their system achieves high accuracies on American Sign Language datasets and the Handicraft-Gesture dataset, outperforming existing approaches. The study highlights the importance of incorporating temporal information from sequential hand gesture data to achieve accurate recognition, enhancing the potential for robust gesture-based interaction systems. 

Manganaro et al. \cite{b18} present the Briareo dataset \cite{b19}, curated specifically for automotive hand gesture recognition. The dataset includes samples captured using various cameras (RGB, infrared stereo, and depth) and 3D hand joints from the LMC. It provides a substantial number of hand gesture samples performed by multiple subjects, enabling the use of deep learning approaches. The authors propose a framework for hand gesture segmentation and classification using this dataset. Their research lays the groundwork for improved gesture recognition systems in automotive environments, enhancing safety and convenience. Boulahia et al. \cite{b20}examine the suitability of an action recognition feature set for modeling dynamic hand gestures using skeleton data from the LMC. Their dataset consists of unsegmented streams of 13 hand gesture classes performed with either a single hand or two hands. The proposed approach surpasses previous methods, contributing significantly to dynamic hand gesture recognition. The study provides valuable insights into feature selection and classification techniques, advancing our understanding of recognizing intricate hand gestures.

Ameur et al. \cite{b21} present a Leap Motion database for hand gesture recognition in medical visualization. Their interactive application utilizes gestural hand control with the Leap Motion Controller, and they propose a tailored 3D dynamic gesture recognition approach for Leap Motion data. The experimental results demonstrate high accuracy in recognizing modeled gestures, contributing significantly to gesture-based interfaces in the medical field and enabling intuitive and precise interactions.

All these studies contribute to the field of hand gesture recognition using the Leap Motion Controller, advancing our understanding and techniques in this area. Through the utilization of temporal dynamics, novel approaches, and curated datasets, researchers are pushing the boundaries of hand gesture recognition and its applications. The versatility and potential of the Leap Motion Controller are highlighted, facilitating seamless interactions between humans and machines. Ongoing research and innovation are driving us towards a future where hand gestures play a crucial role in communication, healthcare, robotics, and more.

\section{Experimental Evaluation}

In this experimental evaluation study, we aim to explore the capabilities of the Leap Motion Controller 2 \cite{b3}, a device that allows users to interact with computers and digital environments using hand and finger gestures. By leveraging advanced motion tracking technology, the controller captures and interprets a wide range of gestures, providing an immersive user experience. Through our experimentation with different gestures, we seek to assess the device's gesture recognition performance and gain valuable insights into its usability and potential applications.

\begin{figure}[h]
\centering
\includegraphics[width=1\linewidth]{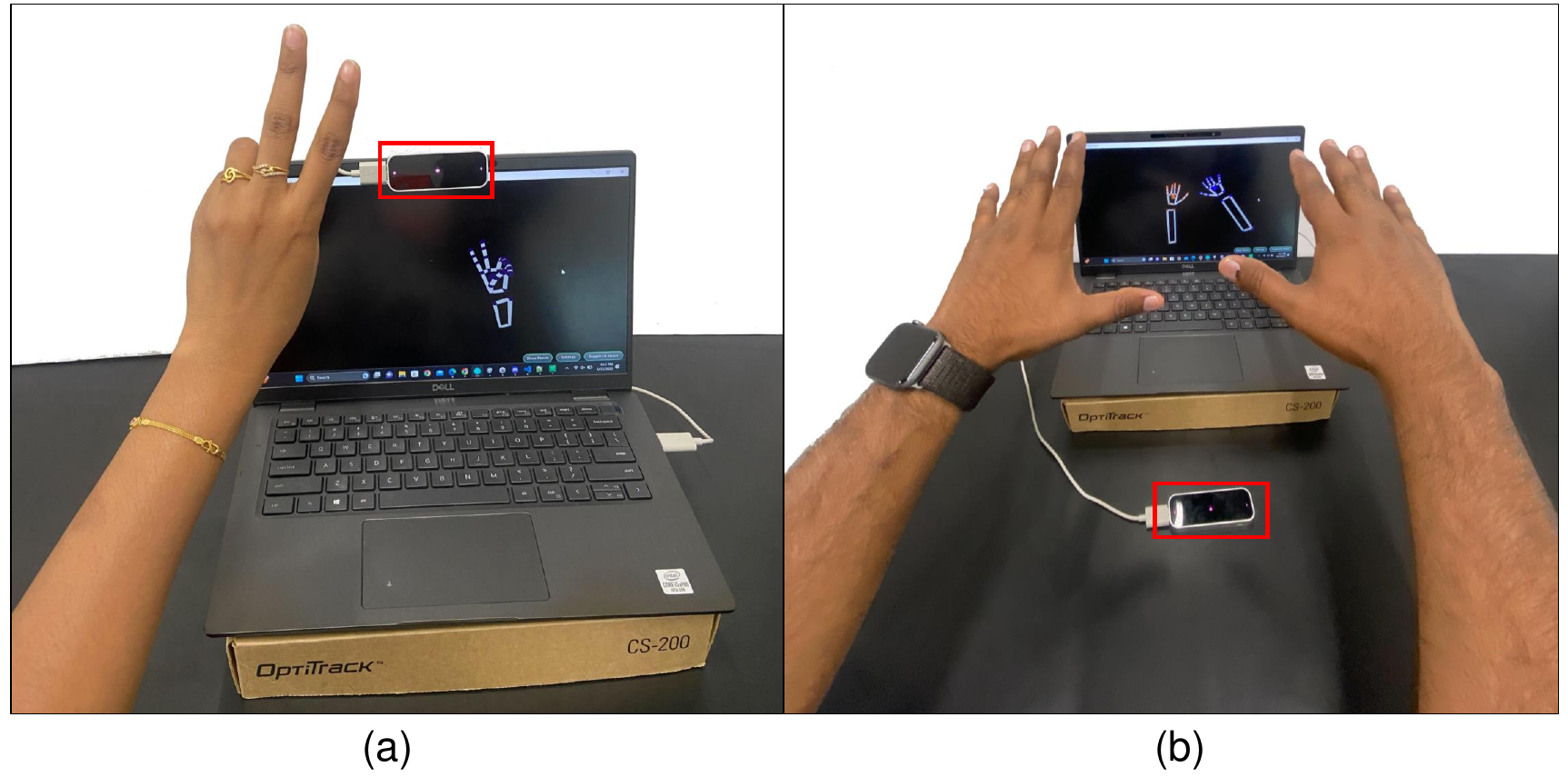}
\caption{The Leap Motion Controller 2 is being utilized in our experimental study, featuring both Screentop and Desktop mode. (highlighted by a red box) (a) depicts a single-hand gesture demonstration, and (b) a two-hand gesture.}
\label{fig03}
\end{figure}

\subsection{Experimental Setup}
In this evaluation study, we utilized a laptop running Windows 10 and Ultraleap's fifth-generation hand-tracking software (v5.12) \cite{b22}. The software module facilitates seamless communication with the Leap Motion Controller 2, while the control panel version (v3.1.0) allows device configuration. The hand-tracking module adaptively and dynamically adjusts frame rates during gesture performance. Our experimentation encompassed different settings, including tracking modes (Desktop and Screentop) and device orientations (Normal and Inverted), as shown in Figure \ref{fig03}. 

\begin{figure}[t]
\centering
\includegraphics[width=1\linewidth]{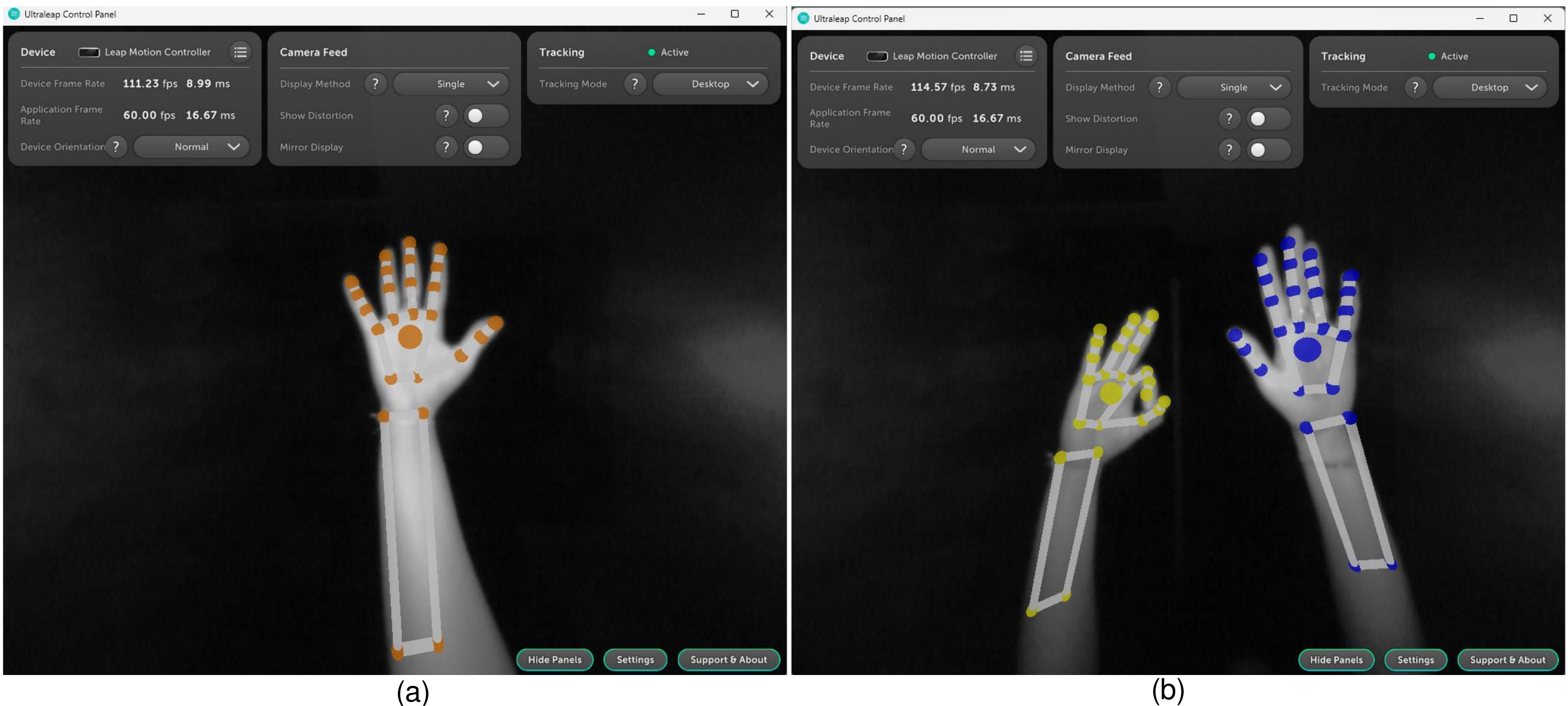}
\caption{Ultraleap Control Panel User Interface. (a) display of single hand gesture and (b) display of both hand gestures, showcasing the stereo images of the respective poses of the hands gestures captured from the Leap Motion Controller 2.}
\label{fig04}
\end{figure}

We also explored display methods such as single camera feed and side-by-side camera feed to optimize gesture visualization. Furthermore, we examined the simultaneous display of stereo images of the hand poses (as in Figure \ref{fig04}) and their reconstructed gestures. Additionally, we inspected the visual effects like raw distorted view and post-processed distortion and the impact of mirror display.
To ensure accurate tracking, we ensured prior calibration using the leap motion control panel's recalibration feature. Achieving a calibration score of 80 or above is crucial for maintaining tracking accuracy. Through this evaluation, we gained valuable insights into the performance, usability, and gesture recognition capabilities of the sensor controller. Through our experiments, we tested the device's ability for fast initialization, tracking diverse hand sizes, supporting interactions with both hands simultaneously, functioning in challenging environments, offering flexible integration options, and showcasing pose accuracy through real-time reconstruction capabilities.

\begin{table}[t]
\centering
\caption{Commonly used gestures across different domains }
\label{table03}
\resizebox{\columnwidth}{!}{%
\begin{tabular}{c|c}
\hline\hline
\textbf{Domain} &
  \textbf{Commonly used Gestures} \\ \hline \hline
Sign Language Recognition {\cite{b23} } &
  Alphabet (A-Z), Numbers (0-9), Greetings, I love you, Thank you, Help \\ \hline
Human-Computer Interaction  {\cite{b24} } &
  \begin{tabular}[c]{@{}c@{}}Pointing, Swiping, Pinching, Grabbing/Grab, Thumbs-up/Thumbs-down, \\ Open/Closed fist, Open/Closed palm, Hand waving\end{tabular} \\ \hline
AR/VR/XR  {\cite{b25}}{\cite{b26}}{\cite{b27}}&
  \begin{tabular}[c]{@{}c@{}}Grabbing/Grab, Pointing, Pinching, Thumbs-up/Thumbs-down, \\ Open/Closed fist, Fist Bump/High-Five, Throwing/Tossing\end{tabular} \\ \hline
Gaming {\cite{b28}} &
  Dual-handed gestures, Fist pumps/Celebration, controller gestures \\ \hline
Human-Robot Interactions {\cite{b29}} &
  \begin{tabular}[c]{@{}c@{}}Follow me, Stop/Pause, handshake, Nodding/Head Shaking, \\ Petting/Rubbing\end{tabular} \\ \hline
Smart Home Technology {\cite{b30}} &
  \begin{tabular}[c]{@{}c@{}}Open/Closed palm, Pointing, Swiping, Thumbs-up/Thumbs-down, \\ Double tap, rotation, increase/decrease, clapping\end{tabular} \\ \hline
Rehabilitation and Therapy {\cite{b31}} &
  \begin{tabular}[c]{@{}c@{}}Fine motor control exercises, grasp, and release, pinch, manipulate, \\ stretch and shrink, squeezing, finger tapping\end{tabular} \\ \hline
Dance and Performing Arts {\cite{b32}} &
  \begin{tabular}[c]{@{}c@{}}Dance-specific hand gestures, Classical hand mudras like Anjali Mudra,\\ Gyan Mudra, and Shuni Mudra\end{tabular} \\ \hline \hline
\end{tabular}%
}
\end{table}

\subsection{Gesture Visualization}

In our experimental study, we conducted a thorough evaluation of Leap Motion Controller 2's capabilities in capturing and visually recognizing commonly used gestures across various domains. To ensure a comprehensive assessment, we focused on domains such as sign language recognition, human-computer interaction, AR/VR/XR, gaming, human-robot interactions, smart home technology, rehabilitation and therapy, and dance and performing arts.
To capture participants' hand movements, we utilized the Leap Motion Controller 2 in conjunction with the control panel visualizer module. The study included a diverse group of participants, comprising both male and female individuals within the age range of 20 to 35 years. Participants were given specific instructions to perform selected gestures from each domain, and visual references were provided to aid their understanding.

\begin{figure}[t]
\centering
\includegraphics[width=1\linewidth]{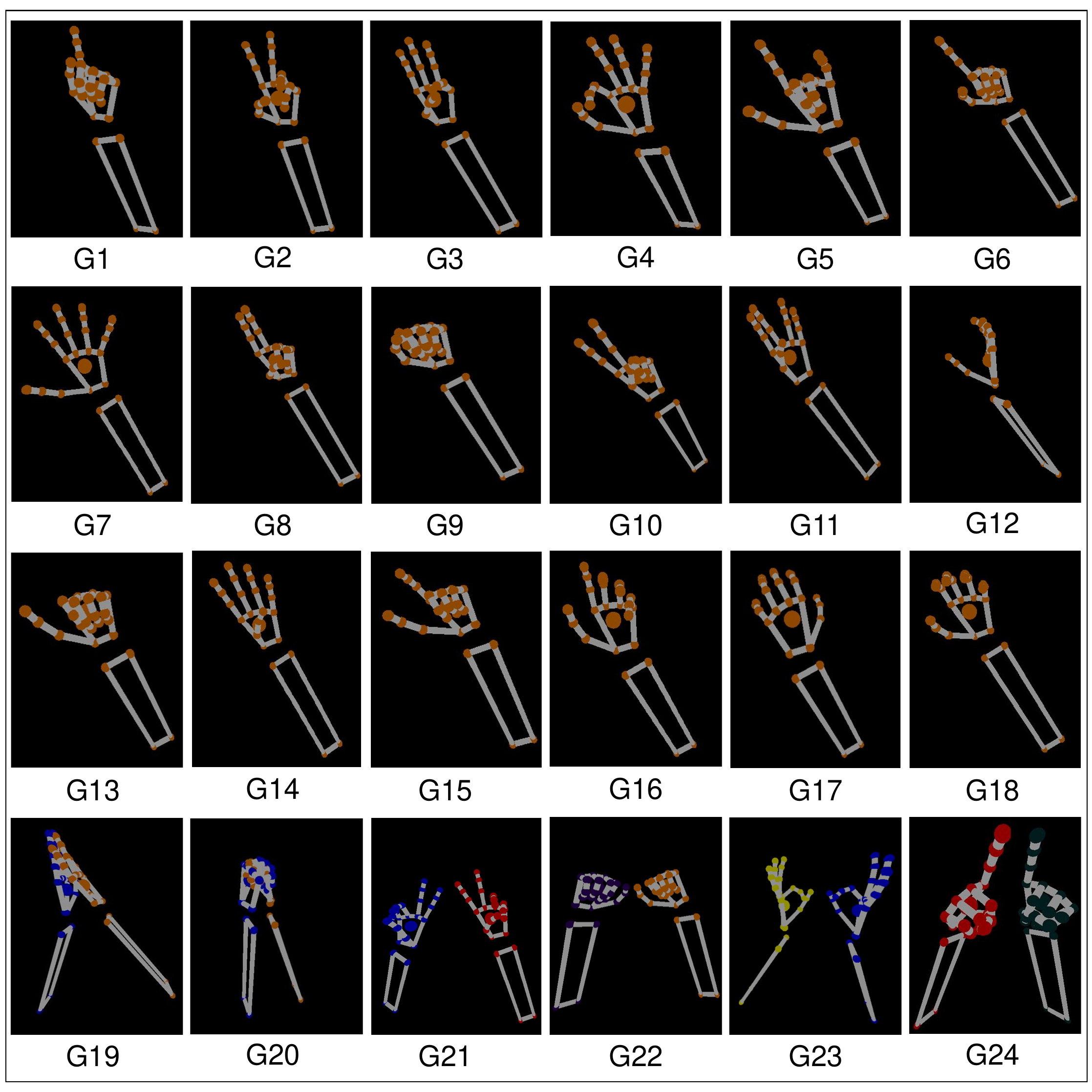}
\caption{Gestures in Various Domains: Each snapshots represents (In rows from Gesture 1 (G1) to Gesture 24 (G24)) gestures performed by participants across different domains, as listed in Table \ref{table03}, illustrating the diverse range of gestures explored in the study.}
\label{fig05}
\end{figure}

Table \ref{table03} summarizes the most commonly used gestures\cite{b23, b24, b25, b26, b27, b28, b29, b30, b31, b32} across the aforementioned domains. Using the real-time visualization capabilities of the Visualizer module, we captured and graphically represented the participants' hand poses as gestures (see Figure \ref{fig05}). This allowed us to visually analyze the accuracy and recognize each gesture. By comparing the performed gestures with the expected gestures from their respective domains, we realized Leap Motion Controller 2's ability to detect and reconstruct a wide range of gestures.

The evaluation of the Leap Motion Controller 2 revealed valuable insights into its performance in capturing and recognizing gestures across different domains. The findings highlight its suitability for a range of applications. For instance, in sign language recognition, the device shows promise in accurately interpreting sign language gestures for assisting the deaf community. In the domain of human-computer interaction, the controller's precise gesture recognition capabilities enhance natural and intuitive interaction with computers. Its integration into AR/VR/XR environments enables users to manipulate virtual objects and enhances immersion. Gesture-based gaming experiences become more intuitive and immersive with the controller. The device also facilitates natural interactions between humans and robots, and its integration with smart home technology adds futuristic gesture control to home automation. In rehabilitation and therapy, therapists can use the controller to guide patients through exercises. Lastly, it holds potential in the realm of dance and performing arts, allowing performers to create expressive and interactive performances through hand movements.

\section*{Conclusion}
This comprehensive review provides valuable insights into the Leap Motion Controller (LMC), an advanced hand-tracking technology that enables touchless interaction and immersive experiences in various domains. The review covered the technical overview of the LMC, including its ability to capture and interpret a wide range of gestures, providing users with an immersive experience. It also highlighted the importance of hand gestures in different domains and how the LMC can effectively support and enhance interactions in these domains. The review further presented a comprehensive overview of existing gesture datasets, showcasing their significance in advancing gesture recognition research. Through an experimental evaluation, the capabilities of the LMC were assessed, considering different settings such as tracking modes, device orientations, and display methods. The evaluation revealed the LMC's impressive performance in real-time gesture visualization, visual pose accuracy, and its ability to support simultaneous interactions with both hands. The findings emphasize the controller's suitability for a wide range of applications accross different domains.

To further advance the field of hand gesture recognition, our future work will focus on creating larger datasets that encompass both static and dynamic gestures. These datasets will enable more robust and comprehensive training of gesture recognition models. Additionally, developing applications in virtual reality (VR) and human-computer interaction (HCI) that leverage the capabilities of the LMC can enhance user experiences and open new possibilities for intuitive and immersive interactions. By continuing to explore and innovate with larger datasets and applications in VR and HCI, we intend to unlock the full potential of hand gesture recognition and create transformative user experiences.

\vspace{12pt}

\end{document}